# Observation of Caustics in the Trajectories of Cold Atoms in a Linear Magnetic Potential

Wilbert Rooijakkers, Saijun Wu, Pierre Striehl, Mukund Vengalattore and Mara Prentiss

*Center for Ultracold Atoms, Harvard University,*
*Physics Department, Lyman Laboratory, Cambridge MA 02138*

We have studied the spatial and temporal dynamics of a cold atom cloud in the conservative force field of a ferromagnetic guide, after laser cooling has been switched off suddenly. We observe outgoing "waves" that correspond to caustics of individual trajectories of trapped atoms. This provides detailed information on the magnetic field, the energy distribution and the spin states.

The transfer of cold atoms from a magneto optical trap (MOT) to a magnetic trap is a crucial step in most experiments on Bose-Einstein Condensation (BEC) in an atomic vapour [1,2]. The efficiency of this transfer has been carefully studied, and the effects of a "plugging" or bias field on the condensation process have also been considered [3]. The plugging field produces a potential that is linear as a function of position at large distances from the center of the trap, but harmonic as a function of position for locations near the center of the trap. There is an intermediate region where the potential is neither linear nor quadratic. In addition, as has been noted in BEC experiments, gravity can play an important role in the motion of the atoms in the magnetic trap [4]. The trajectories of the atoms in the linear magnetic trap and the effects of a plugging field on those trajectories have not yet been considered in detail.

In this paper, we demonstrate both theoretically and experimentally that the non-harmonic nature of the linear potential can result in caustics in the trajectories of the atoms where the atom densities in the trap are highly non-uniform and characterized by discrete ring like structures. This is the first observation of these structures in cold atom trajectories, though they had been predicted to occur in atom lithography experiments where the time of the experiment was longer than 1/4 of the oscillation period characterizing the inner harmonic portion of the wells [5, 6]. We report on the influence of gravity and plugging fields for a horizontal magnetic guide and show that theory and experiment are in good agreement.

Our experimental setup has been described elsewhere in great detail [7]. Using long mu-metal coils underneath a gold mirror (Fig. 1) we generate a magnetic field that can be described by:

$$B_x = B'y\vec{e}_x$$
$$B_y = -B'x\vec{e}_y \quad (1)$$
$$B_z = 0$$

with B'=dB/dr the magnetic field gradient. In the absence of any laser light but in the presence of a plugging field $B_z=B_{plug}$ the atomic motion can be thought of as governed by a conservative potential:

$$V(r) = V_{magnetic}(x, y) + V_{gravity}(x) = M_F g_L \mu_B \sqrt{B_{plug}^2 + B'^2(x^2 + y^2)} + mgx \quad (2)$$

For typical magnetic field gradients B' used in the experiment (25 G/cm < B' < 60 G/cm), gravity cannot be neglected and results in one dimensional distortion of the cylindrically symmetric potential due to the magnetic field gradient. For $^{87}$Rb, F=2, $M_F$=2 the effect of gravity is equivalent to a Stern-Gerlach force in a gradient B'=$m_{Rb}g/\mu_B$=15 G/cm, with g=9.81 m/s$^2$ the gravitational acceleration and $m_{Rb}$ the atomic mass. This is the minimum field gradient required to hold the atoms in the magnetic waveguide.

In our experiments we prepare $^{87}$Rb in the F=2 ground state, with a Landé factor $g_L$=1/2 and $m_F$ can take the values $M_F$=-2,-1,0,1,2. Non-adiabatic transitions (between $M_F$ levels [8]) could occur in regions of weak field (notably the origin {x,y}={0,0}), but are suppressed due to poor overlap of the atom cloud with the origin, and can be forcibly suppressed by applying a plugging field, which lifts the magnetic degeneracy everywhere. We shall demonstrate that there is no evidence of any non-adiabatic transitions in our experiment. In the absence of a plugging field the magnetic potential is radially symmetric and linear: $V_{magnetic}(r) = M_F g_L \mu_B B' |r|$.

Atoms can be loaded into this magnetic waveguide using a 2D surface MOT [7], with laser beams along the x = y and x = -y directions for radial confinement and along the z-axis for longitudinal damping. The radial confinement is well described by MOT theory [7] using $\sigma^+$-$\sigma^-$ polarization for the counterpropagating laser beams. We find experimentally that any polarization can be used along the z-direction. This indicates that along this axis optical molasses theory can be used. In the present experiment we use $\sigma^+$-$\sigma^+$ polarization along the z-axis, providing natural on-axis optical pumping in cases where a plugging field $B_z = B_{plug}$ is applied.

Before the dissipative light is turned off, the atoms are in a non-conservative quasi-potential that is quadratic in the radial distance r (in the limit when the magneto-optical trapping force can be described by $F_{MOT} = -\kappa r$, where $\kappa$ is the spring constant). The MOT light is switched off abruptly using an acousto-optic modulator (switching time measured to be < 5μs), backed up by a mechanical shutter. This ensures that the atoms retain their velocity, but find themselves in a new conservative potential given by eq. (2). After the transfer to the new potential three situations may occur: (i) the average kinetic energy at t=0 is much smaller than the average potential energy at t=0 (Fig 2a), (ii) they are of equal order (Fig. 2b), or (iii) it is much larger (Fig 2c). The second case (ii) has been pursued by many as a matching (phase space overlap) condition to effectively transfer atoms from a 3D MOT to a 3D magnetostatic trap, before engaging in evaporative cooling to achieve Bose Einstein Condensation [2]. Such matching can be achieved by ramping up the magnetic field gradient at the same time or slightly before the MOT light is switched off, and depends also on the intensity and detuning of the MOT light. In our experiments the field gradient required for matching the size of the magneto-optically trapped cloud with that of the magnetically trapped cloud was approximately 80 G/cm for $^{87}$Rb or $^{133}$Cs [7].

We have made a numerical analysis of the atomic trajectories in a one-dimensional (1D) potential analysis, as shown in Fig. 2. From Fig. 2 it is observed that in cases (i) and (iii) caustics (envelope functions) develop, related to the turning points of individual trajectories with different total energies in the linear potential. These caustics result in an apparently outgoing wave or halo, which seems to propagate with near constant speed. The halo is consequence of the fact that particles catch up with one another and pass each other in the anharmonic potential.

An atom moving in a 1D linear confining potential is subject to a constant acceleration a, which changes sign as it traverses the origin. It is easy to work out that the time t for n oscillation periods T with amplitude A is given by:

$$t = nT = 4n\sqrt{\frac{2A}{a}} \tag{3}$$

Taking the derivative of A with respect to time t provides an approximate expression for the "velocity" of the caustic wavefront:

$$v_{group}(t) = \frac{at}{16n^2} \tag{4}$$

For a cloud that is initially squeezed in momentum (Fig 2a) the first caustic appears for n=½; for a cloud initially squeezed in position (Fig. 2c) it appears for n=¼. A new caustic appears for every increase of n by ½. Thus halos will keep appearing, as is observed in Fig 5a. This model is only true for a symmetric linear potential, but can easily be extended to an asymmetric potential in the presence of gravity (Fig. 4).

It is important to note that the caustics are a consequence of the anharmonicity of the potential Eq. (2). To illustrate this we have plotted the trajectories for different values of $B_{plug}$ in Fig. 3. Increasing $B_{plug}$ makes the potential more harmonic near the center of the potential, such that atoms with small energy oscillate with equal frequency. Furthermore it is observed that with equal initial kinetic energy atoms remain more confined in the linear potential when $B_{plug}$ is reduced.

So far we have discussed the caustics in 1D potentials. We have made a full 3D classical trajectory simulation to make sure the model carries over to more dimensions. Depicted in Fig. 6 are a numerically produced series of images, that show an initial expansion, a falling droplet and several halos that appear as multiple outgoing "waves" when played as a movie.

We now turn to the experiment: after we switch off the MOT light that prepares a cold atom cloud from the background vapour (vapour pressure ~ $10^{-9}$ mbar), absorption images *along* the cloud are taken as shown in Figure 1, exploiting the long direction for large absorption. A double achromatic lens (f=30 cm) with a diameter of 5 cm at a distance of 60 cm from the cloud images the cloud onto a CCD camera with a resolution of 7 μm/pixel. The camera is shuttered electronically to open only during the time in which the probe beam is flashed. This exposure time was typically 80 μs for absorption images, but could have been reduced to 5 μs if the speed of the atom dynamics had been faster.

Figure 7 shows a series of absorption images taken with a gradient of 24 G/cm, in steps of 1 ms after MOT dissipation was switched off. The initial absorption images in our experiment suggest a density of 5x $10^{11}$ cm$^{-3}$, corresponding to a maximum collision rate of 50 s$^{-1}$, not enough to explain dynamics on the 1 ms timescale, and ruling out collisions playing an important role. This figure is characteristic for the behaviour at other field gradients and does not change when a weak plugging field is applied. In the first 5 ms a rapid expansion can be observed. This expansion is due to the initial kinetic energy or due to strong-field seeking states $M_F$=-1,-2 which are accelerated away from the center of the guide. We have plotted the full width half maximum (FWHM) of the cloud in these first moments in Figure 8. After 5 ms we observe a dense core with a maximum size (FWHM) of 200 μm. Later in time the core breaks up with another core falling out and a halo surrounding the two cores [9].

As different atoms reach their turning point in the potential at different times an absorption measurement of the atomic density at those turning points could provide the full energy distribution inside the magnetic guide. The caustics themselves are only a function of the field gradient and will appear at the same place at the same time independent of the initial temperature, thus mapping out the distribution of initial kinetic energy (that is: if they are visible at all, and not smeared out as in Fig. 2b).

In Figure 8 we have plotted the position of the cloud that falls out versus time, as well as the horizontal width of the halo versus time. The position of the falling cloud can be fitted to accelerated expansion x=½at$^2$, and we find a=4.4 m/s$^2$. Clearly these are not $M_F$=0 atoms which would fall with a=9.8 m/s$^2$. If we assume that these are $M_F$=1 atoms, then the magnetic field gradient is 2*(9.8-4.4)/9.8 * 15 G/cm=16.5 G/cm, much different from the calibration result. The explanation is that we are not observing the same atoms falling but *different*, bound, $M_F$=2 atoms reaching their outer turning point at different times, with a caustic producing local density enhancement. From a fit of the width of the $M_F$=2 caustic we find a gradient of 27 G/cm, which is within the experimental uncertainty of the calibration result.

The presence of multiple halos are due to higher order caustics and we have observed them experimentally (Figs. 9 and 10). Even a third and fourth halo have been detected, as can be understood from the simple model calculations in Fig 5. For the gradient chosen in Fig. 8, gravity cannot be neglected and provides a more significant pull on the $M_F$=1 atoms, as compared to $M_F$=2. This causes the "droplet" shape of the $M_F$=1 atoms, as verified by the numerical simulation in Fig. 6, where the initial $M_F$ state can be chosen *a priori*.

The local density at the caustic is sufficiently enhanced that we measure its position versus time accurately. Figure 10 shows a series of measurement at different values of the magnetic field gradient. The diameter of a caustic ring was measured in the horizontal dimension, perpendicular to the direction of gravity. The caustics appear regularly, e.g. an atom with amplitude 300 μm in a gradient of 40 G/cm reaches its turning point (as measured by the caustic position) first after 9 ms, then after 17 ms, and then 25 ms. This corresponds well to the calculated half oscillation period (Eq. 3) of T/2=9.5 ms (a=26.2 m/s$^2$ for a gradient of 40 G/cm). Conversely: a precise measurement of this period, which is not convolved by the temperature nor the initial width of the initial cloud could lead to a more accurate calibration of the magnetic field gradient. The data in Fig. 10, combined with additional data at 20 G/cm, also show that the oscillation frequency scales with the inverse square root of the applied field gradient.

An effect that may modify the observed time dynamics are Majorana transitions. These are non-adiabatic transitions between $M_F$-states, converting positive $M_F$ states (weak field seekers) into negative $M_F$ states (strong field seekers). In previous experiments in 3D quadrupole traps Majorana flips were a serious impediment for achieving higher phase space densities [1]. The transitions would occur continuously as the atoms "probe" the magnetic center, where they are unable to adiabatically follow the preferred field direction. The spin flip rate should dramatically change when a plugging field is applied, lifting the degeneracy and reducing the flip rate. We have applied up to 0.4 Gauss (corresponding to a Larmor frequency $\omega_L$>560 kHz for $M_F$=2 atoms and $\omega_L$>260 kHz for $M_F$=1, whereas most trap oscillation frequencies are under 1 KHz) and see no change, ruling out non-adiabatic transitions.

Although the experimental results are fully described by the simple theory outlined above, we shall now discuss further refinements to the theory that may be necessary for higher atomic densities and smaller temperatures, up into the quantum degeneracy regime. It is instructive to use a wave mechanical picture instead of classical trajectories. In most general form the equation of motion of atoms in the magnetic field is given by a non-linear Schrödinger equation (Gross-Pitaevski approximation) for an F=2 spinor, consisting of 5 $M_F$ states:

$$(-\frac{\hbar^2}{2m}\nabla^2 + mgz + \mu_B B_x \vec{J}_x + \mu_B B_y \vec{J}_y + \mu_B B_z \vec{J}_z + \sum_{k,l} g_{kl} \Psi_k^* \Psi_l) \begin{pmatrix} \Psi_{-2} \\ \Psi_{-1} \\ \Psi_0 \\ \Psi_{+1} \\ \Psi_{+2} \end{pmatrix} = \frac{\hbar}{i}\frac{\partial}{\partial t} \begin{pmatrix} \Psi_{-2} \\ \Psi_{-1} \\ \Psi_0 \\ \Psi_{+1} \\ \Psi_{+2} \end{pmatrix} \quad (5)$$

The first and second term on the left hand side of this equation are diagonal and describe kinetic and gravitational energy respectively. The other terms may provide coupling between the sublevels: the part with the five-by-five spin matrices $J_x$, $J_y$ and $J_z$ describes the Zeeman interaction and the last term describes non-linear interaction due to collisions, in the mean field approximation characterized by a coupling strength $g_{kl}$, with indices k and l denoting the $M_F$ state. As we have outlined before there is no coupling observed in the experiment, rendering system (5) into a set of 5 independent and linear Schrödinger equations. However the study of spinor dynamics is an area of active theoretical [10-14] and experimental [15] research, and is relevant for future applications of the magnetic waveguide described in this paper. Very recently the F=1 spinor equivalent of equation 5 has been solved in the presence of a Ioffe-Pritchard magnetic field [27].

In Figure 5 we have plotted the solution of the 1D Schrödinger equation for the $M_F$=2 state in the combined magnetic and gravitational potentials (2), together with the classical trajectories picture. Clearly the caustics are observed in the wave mechanical model as well: regions of increased probability for the quantum wavefunction. Correspondence of the two models has been noted by [16] and coined the term 'quantum carpets' by [17, 18] when these regions of enhanced probability are due to mode interference. The situation described in this paper can be understood completely classically and it is not necessary to invoke quantum interference.

For our geometry with a potential V(r)=F|r| the density of states is given by [19]:

$$\rho(\varepsilon) = \frac{2\pi^2 m}{h^2}\left(\frac{\varepsilon}{F}\right)^2 \tag{6}$$

Integrating this over the energy range observed in the experiment, we find that ~$10^6$ modes are populated, enough to find perfect agreement between classical and quantum dynamics. For lesser modes significant deviations can be found as pointed out by Gea-Banacloche [20], who examined a quantum bouncing ball in a gravitational field en Wallis and coworkers[21], who investigated a gravitational cavity. Individual modes, trapped in a linear potential provided by gravity, have been observed with neutrons [22].

Finally we discuss a diffraction experiment to measure the initial size $\sigma_x$ of the cloud, which is not revealed in the previous absorption data of Fig. 7: the optical density $OD_{//}$ along the cloud is too large: $OD_{//}$=100 for t=0. Thus an absorption measurement with a resonant probe beam as depicted in Fig. 1 is fully absorbed in the center, resulting in an clipped profile. We have imaged the cloud from the transverse side (Figure 11), where the optical density $OD_\perp$ is much less [23]. Using the same magnification we observe a different profile in absorption and fluorescence. The imaging lenses are focused using the fluorescence signal, and their position is not changed when absorption images are taken. There is a clear difference between 2f:2f absorption imaging (with one lens) and f:2f:f absorption imaging (with two equal lenses), due to the phase coherence of the probe light. Only in the latter case the phase plane of the incident probe beam is reconstructed in the image plane, resulting in a crisper image.

Probe light diffraction and refraction by the long and narrow cloud is not negligible, and even plays a role in the transverse direction. The cloud can be described as a medium with a complex index of refraction [24,25]:

$$n = n' + in'' = 1 + \frac{3n_{At}\lambda^3}{8\pi^2}\frac{i - 2\delta/\Gamma}{1 + (2\delta/\Gamma)^2} \tag{7}$$

where λ, δ and Γ are the wavelength of the probe light, the detuning and the linewidth of the atomic transition respectively. The imaginary part n" gives rise to absorption, the real part n' gives rise to phase shift of order $\Phi=OD_\perp |n'-1|/\lambda$. The index of refraction varies strongly near resonance, and therefore absorption near resonance does not provide a good measurement of the size of the cloud. From the fluorescence however we find a sharply spiked distribution (Fig. 11) that fits well to a Lorentzian shape: $\frac{C}{1+(x/\sigma_x)^2}$ with $\sigma_x$=21 µm. Note that the fluorescence measurement does not represent the atomic radial distribution, but its integral along the line of sight. To study further the phenomenon of refraction by the cloud we took out the imaging lenses and detuned the probe laser. The far field pattern at distance R was recorded on the CCD camera, with results plotted in Fig. 12. The observed intensity pattern has been fitted to:

$$I(\rho) - I_0 = I_0 \cos[\phi - \phi_0 + \frac{k\rho^2}{2}\left(\frac{1}{R} - \frac{1}{R_p}\right)]\Im(\rho) \qquad (8)$$

where R=50 cm is the imaging distance (distance between the cloud and the camera) and $R_p$ the radius of curvature of the probe laser. The "envelope" function $\Im(\rho)$ is inversely proportional to the original cloud width, and is related to the fact that at infinite distance the Kirchhoff diffraction integral equals a Fourier transform of the spatial distibution in the object plane. For our purpose we have fitted an envelope function $\exp(-\rho^2/\sigma_\rho^2)$ for which we find $\sigma_\rho$= 2.5 mm, corresponding to a scale size $\sigma_x=R\lambda/\sigma_\rho$=150 µm of the cold atom cloud. However, the observed radial distribution is distinctively non-Gaussian and presents a subject for future investigation. The fitted values for the phase shift are $\phi_0$= -0.4, -1.5 and –2.0 rad for $\delta$=0, -$\Gamma$, -2$\Gamma$ respectively, which confirms that the atomic cloud has $OD_\perp$ ~1.

In conclusion we have presented experiments and numerical simulations to examine the spatial and temporal dynamics of atoms in the linear potential of a magnetic waveguide. This provided detailed information on the magnetic field, the energy distribution and the spin states, which can be used for future applications such as multimode atom interferometry using ferromagnetic guides. In-situ loading of a magnetic waveguide has also been proposed as a mechanism to seed a continuous atom laser [26]. In that case the "mapping" from the dissipative system to the conservative potential is provided by an optical pumping mechanism rather than switching in time, but we expect similar dynamics as described in this paper.

We thank Richard Conroy for his help with building up the experiment and gratefully acknowledge discussions with Gary Zabow, Dr. Michael Moore, Prof. Erik Heller and Prof. Misha Lukin. This work is supported by MURI grant No.Y-00-0007

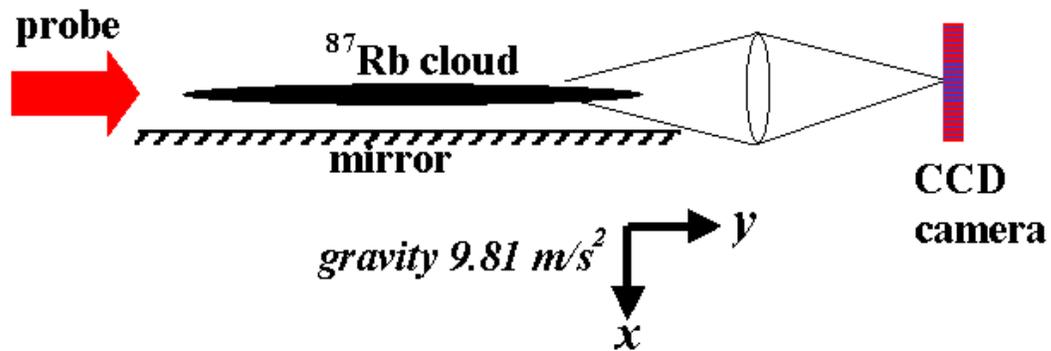

Fig 1: Experimental setup: a laser cooled $^{87}$Rb cloud (length ~5 mm) is prepared at a height of 3 mm above a gold mirror using a surface magneto-optical trapping technique. After switching off the trapping laser beams (not shown), an absorption image (2f:2f imaging with focal length f=30 cm) is taken in the long direction of the cloud using a probe laser beam with a waist of 1 cm.

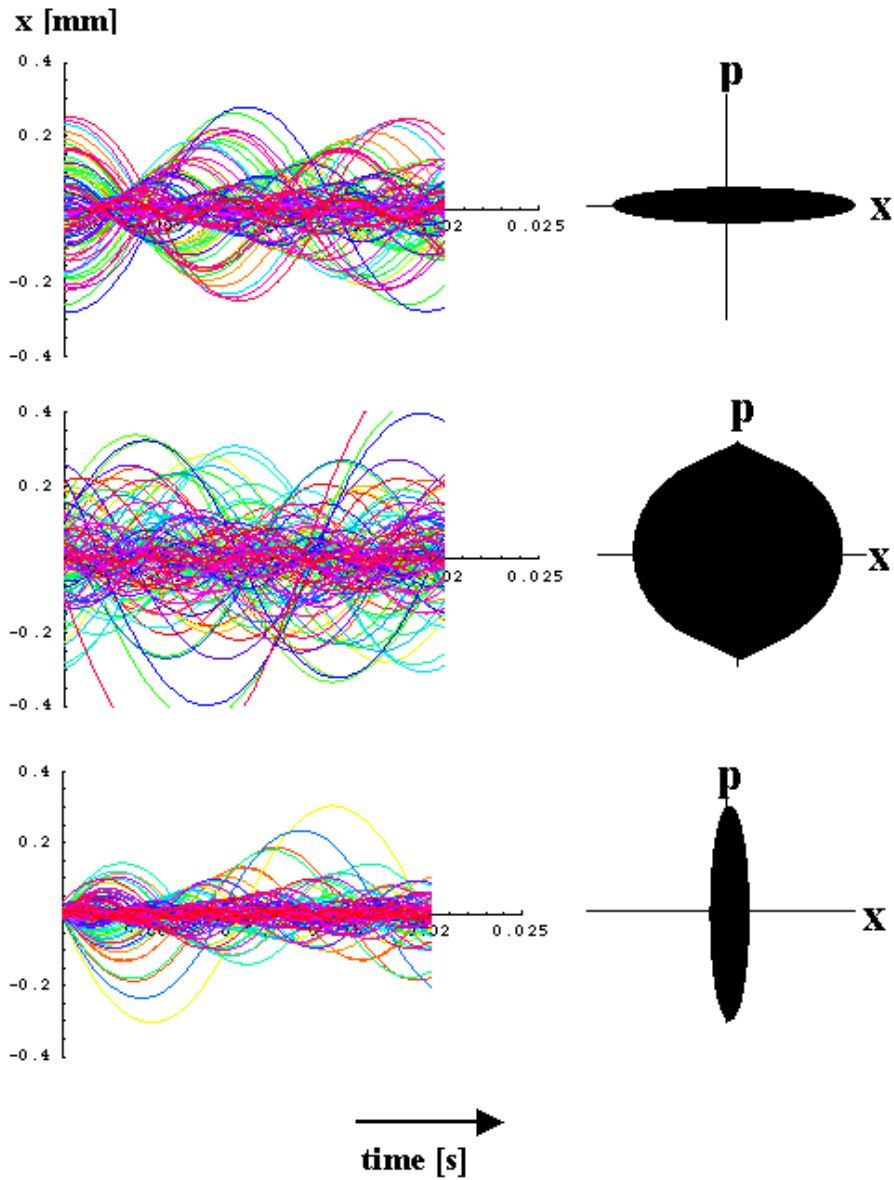

Figure 2: Manifold of 120 trajectories in a linear 1D confining potential (magnetic field gradient 38 G/cm, no gravity) for different conditions of the initial position and momentum distributions. Top: squeezed in momentum: $\sigma_v$=6 mm/s, $\sigma_x$=100 μm, middle: not-squeezed: $\sigma_v$=4 cm/s, $\sigma_x$=100 μm, bottom: squeezed in position $\sigma_v$=4 cm/s, $\sigma_x$=10μm. Caustics are clearly visible in the top and bottom graphs, and correspond to the turning point of individual trajectories. The term 'squeezing' refers to matching of the relevant energy scales before and after transfer to the conservative potential (see text).

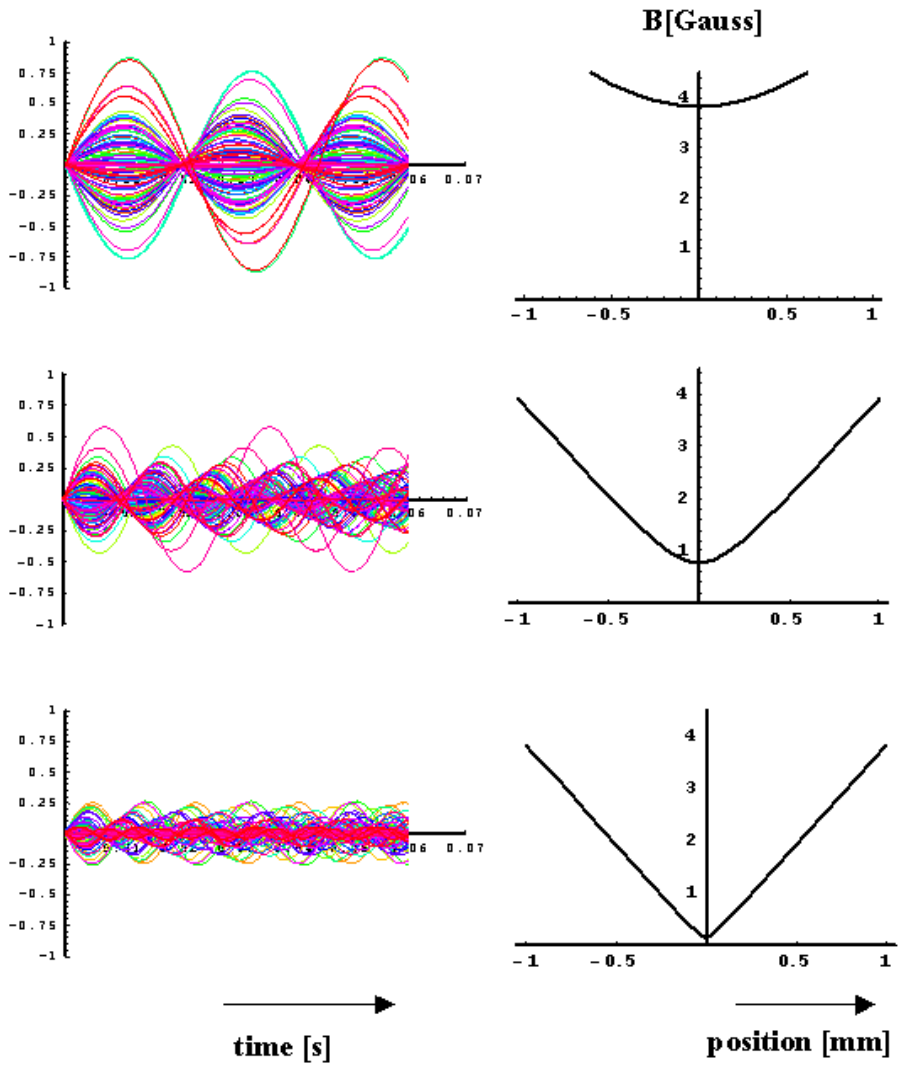

Figure 3: Changing the harmonicity. Left column: manifold of trajectories. Right column: corresponding 1D potential with dB/dr=38 G/cm and $B_{plug}$=3.8, 0.8 and 0.2 G from top to bottom. Note that with the same initial kinetic energy ($\sigma_v$=4 cm/s) the cloud occupies a much greater space for a (nearly) harmonic potential.

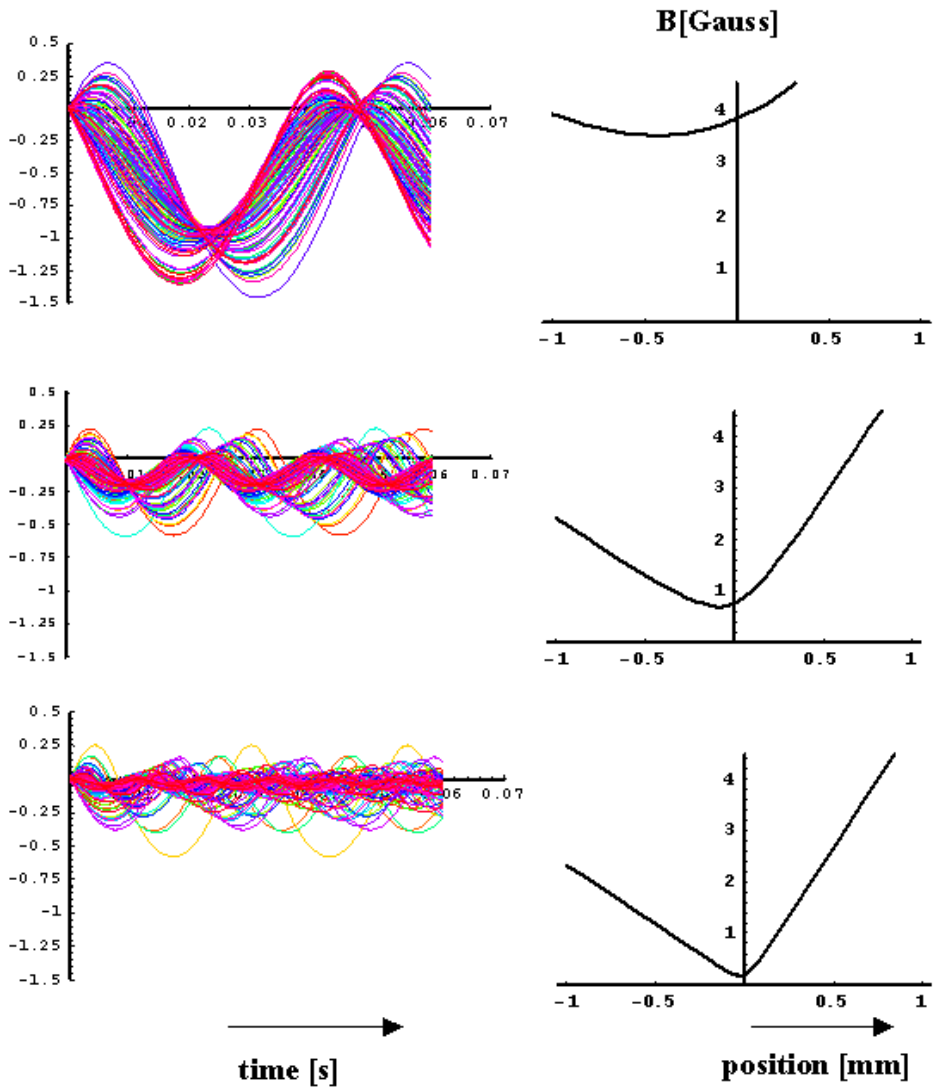

Figure 4: Including the effect of gravity, which corresponds to an extra vertical magnetic gradient of 15 G/cm, in addition to the gradient of 38 G/cm. Same parameters as Figure 3. Note that the potential minimum is no longer at the magnetic centre, and shifts more for larger $B_{plug}$. Caustics with different slope can still be observed for a linear potential.

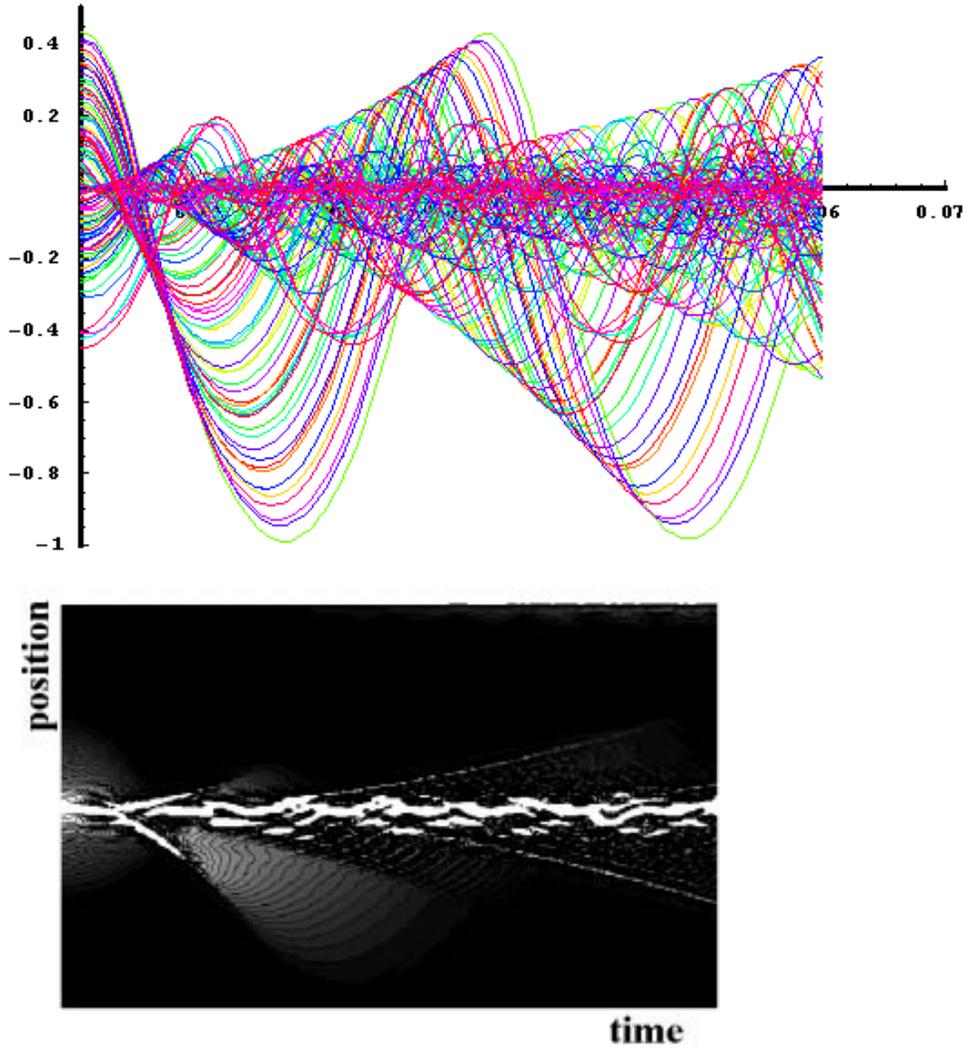

Figure 5: Comparison of 1D classical trajectories (top) and a 1D quantum mechanical model (bottom). Results for the quantum mechanical model are obtained by time propagating an initial wavepacket $\Psi=\exp[-x^2/\sigma_x^2]$, with homogeneous phase, using the time dependent Schrödinger equation. Magnetic field gradient 23 Gauss/cm.

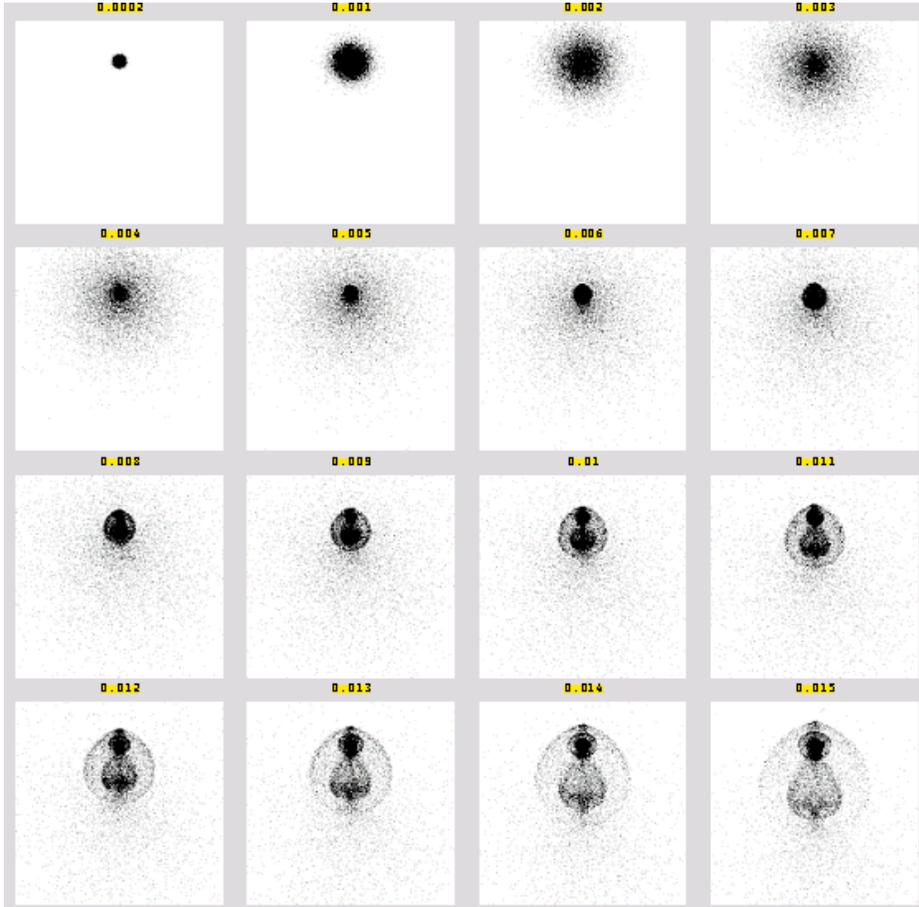

Fig. 6: Numerical images (times as indicated) based on 12000 classical 3D trajectories with 50% of atoms in the $M_F=1$ and 50% in the $M_F=2$ state, dB/dr= 41 G/cm, $\sigma_x=10$ μm and $\sigma_v=7$ mm/s. The atoms are unbound in the long direction of the waveguide. The bottom "droplet" are the $M_F=1$ atoms. Grayscale (inverted) indicates number of atoms in a 10 μm x 10 μm bin. Frame size: 1.5 mm x 1.5 mm.

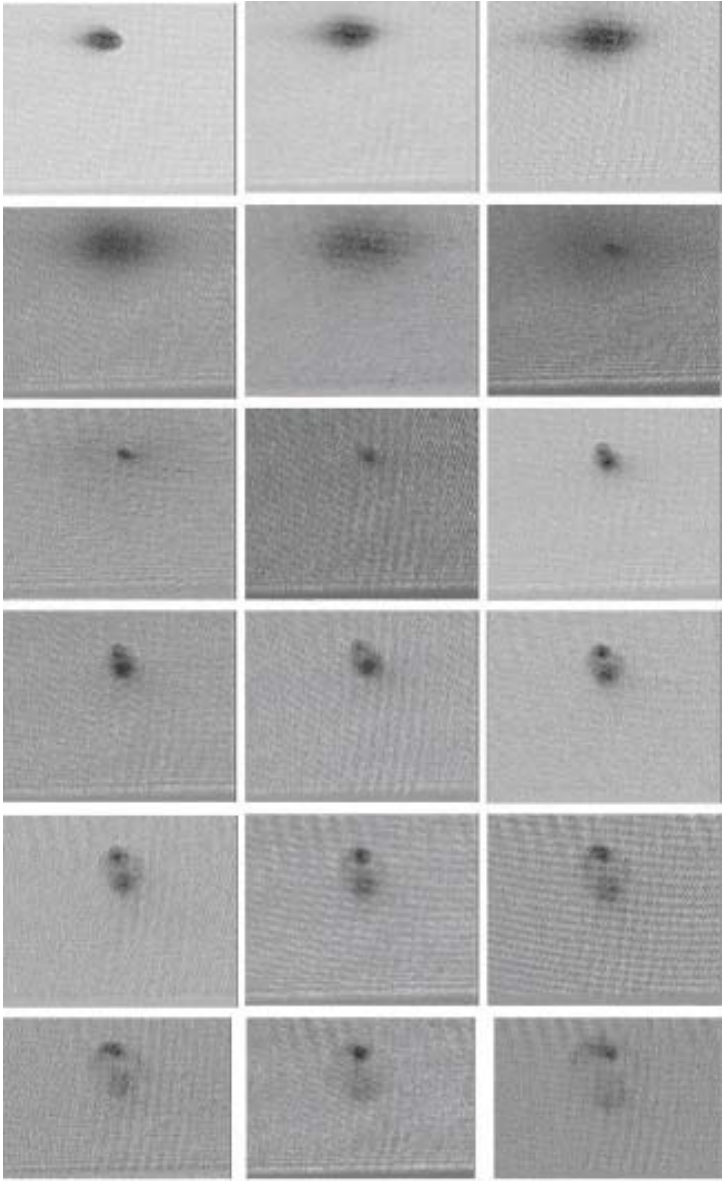

Fig 7: Time-evolution of the cold atom cloud in the magnetic field of the guide (gradient 24 G/cm). Absorption images with resonant coherent probe light in the long direction of the cloud in time steps of 1 ms after the MOT light is switched off. Frame size: 3.5 mm x 4.2 mm. Exposure time 80 μs.

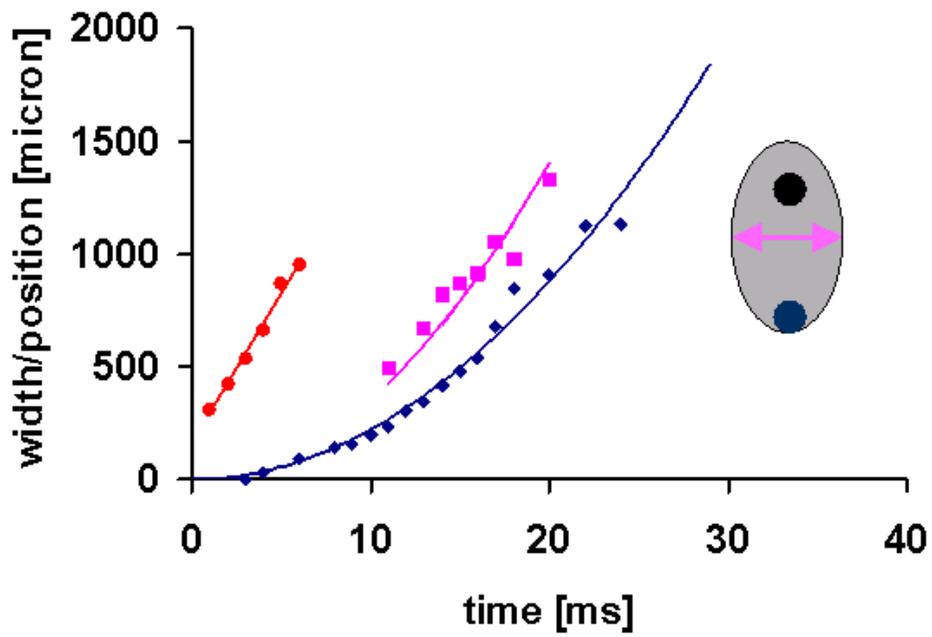

Figure 8: Quantitative plot of position information in Fig. 2. Red ●: FWHM of the cloud in the initial rapid expansion phase. Solid line is a linear fit to the data. Blue ◆: center position of the falling $M_F=1$ cloud. Solid line is fit to find vertical acceleration a=4.4 m/s$^2$, Pink ■: maximum width of the halo in the horizontal direction (not distorted by gravity). Solid line is a fit to Eq. 4.

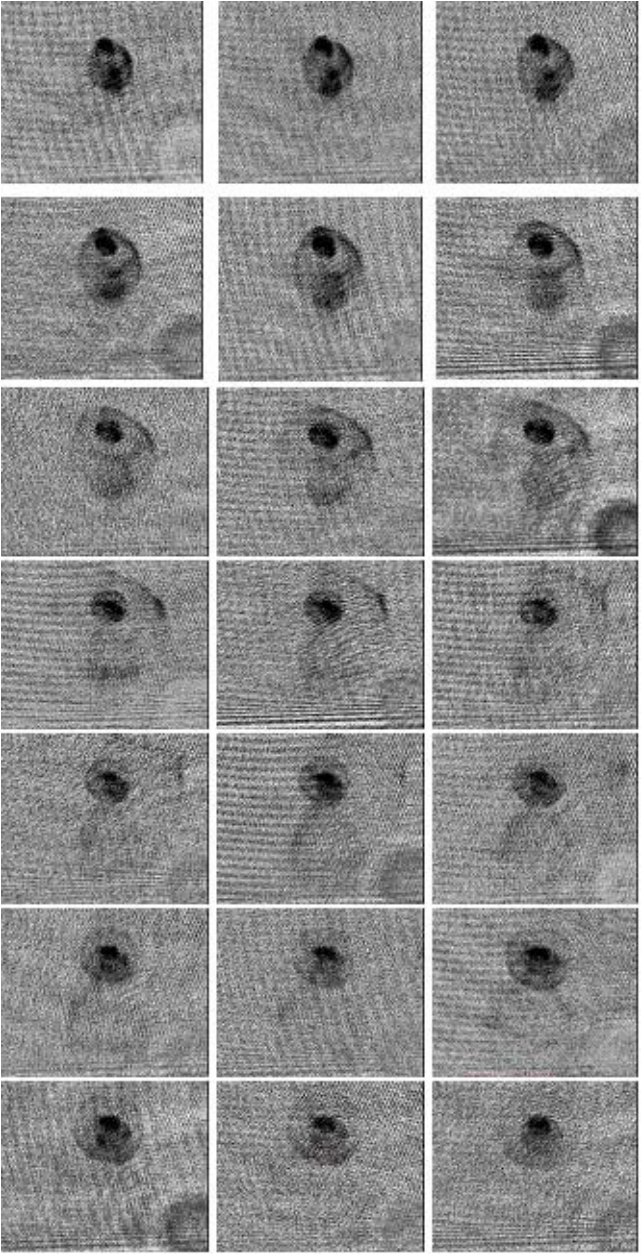

Figure 9: Time evolution of the cold atom cloud (gradient 30 G/cm). Frame size 2.8 mm x 3.5 mm. Same situation in Fig. 7, but we start at t=15 ms, with time steps of 1 ms. An outgoing caustic for $m_F=2$ atoms has emerged at t=15 ms, followed by another starting at t~23 ms. The scarcely visible "droplet" at the bottom corresponds to the more weakly confined $M_F=1$ atoms.

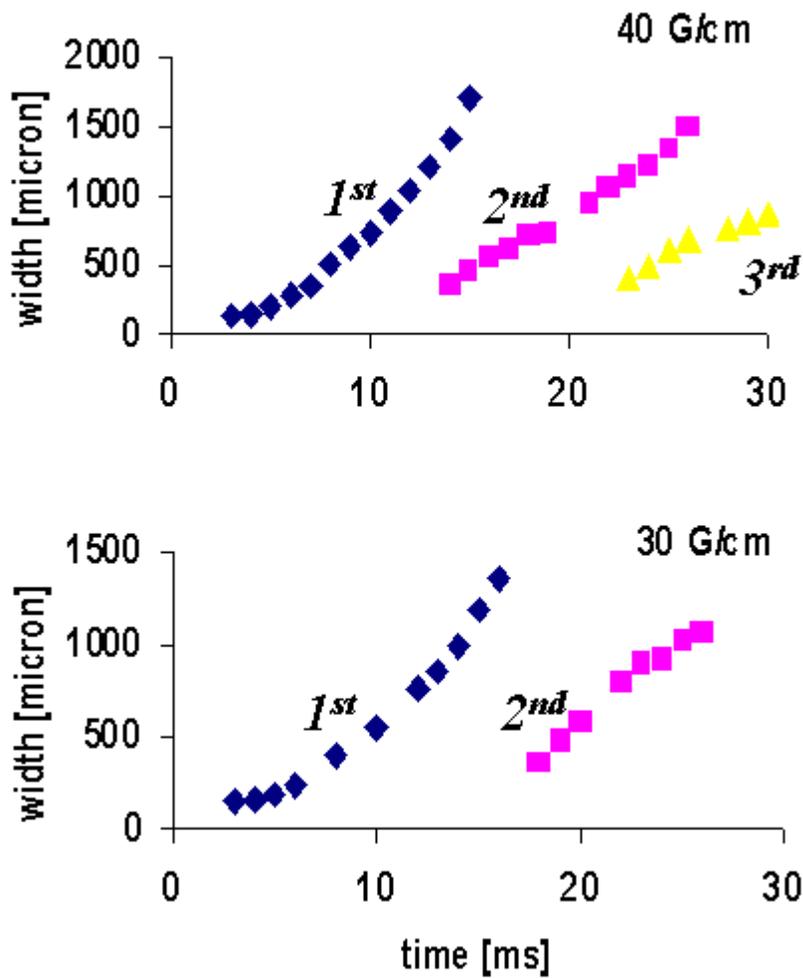

Figure 10: Horizontal diameter of the expanding caustic versus time. Top: magnetic field gradient 40 G/cm, bottom: 30 G/cm. Multiple caustics (labeled 1st, 2nd, 3rd as the order in which they first appear) may be visible at the same time, since the oscillation frequency is a function of amplitude for an anharmonic potential.

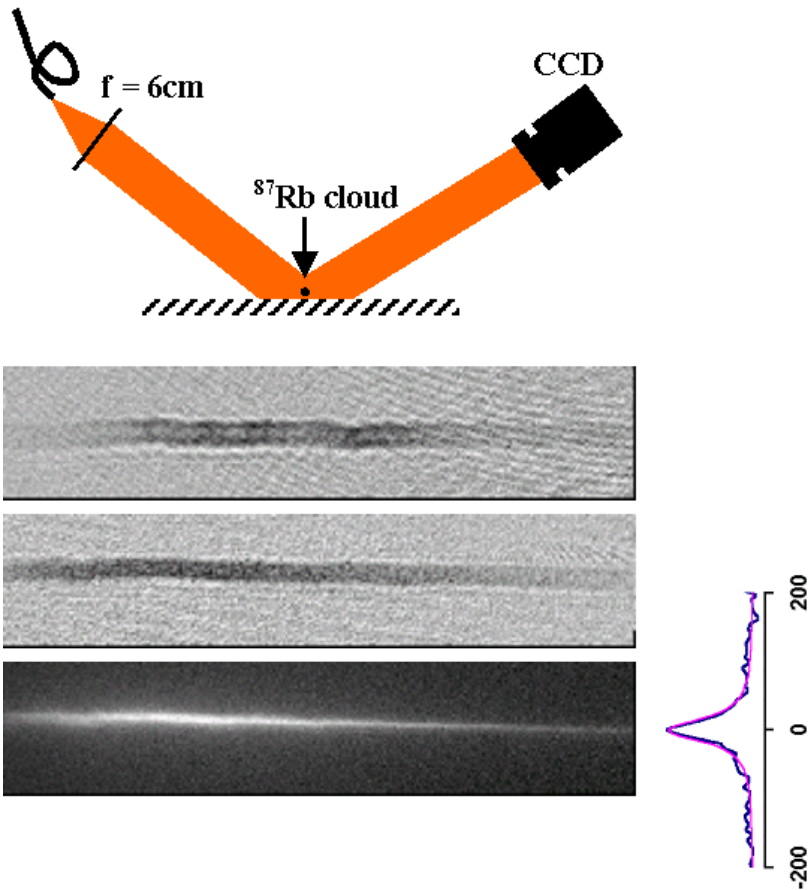

Fig. 11: From top to bottom: a) setup, b) absorption with 2f:2f imaging (f=10 cm), c) absorption with f:2f:f imaging, d) fluorescence image with 2f:2f imaging and view of the cross section profile (units: µm), fitted to a Lorentz (see text). Frame size: 3.8 x 0.8 mm.

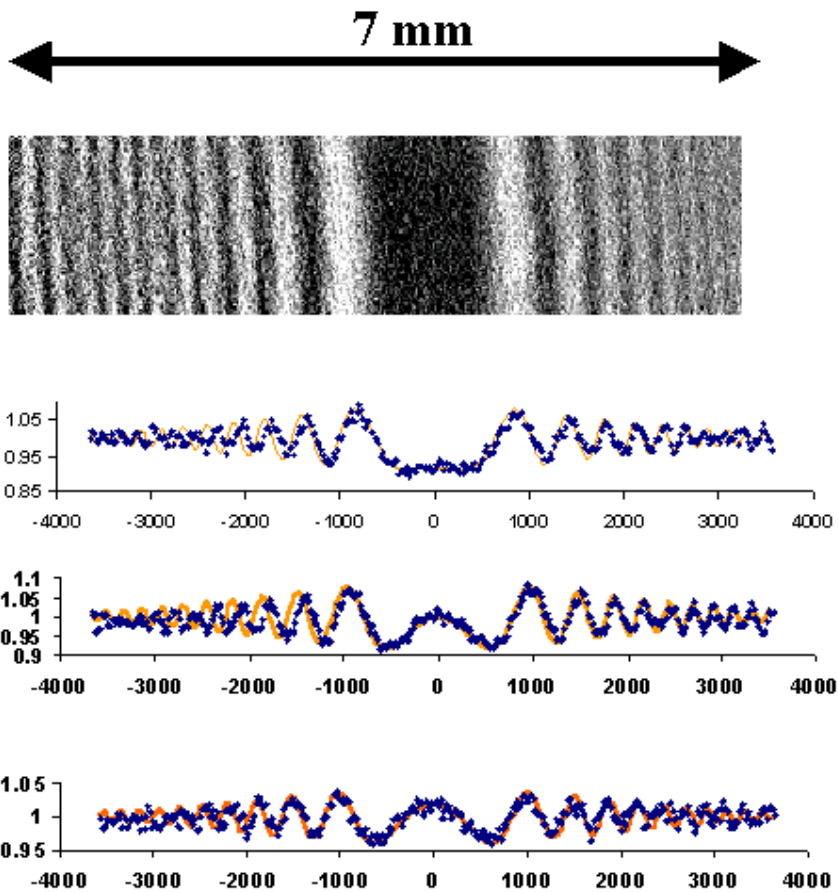

Fig. 12: a) diffraction image of coherently scattered probe light for probe detuning δ=0. Intensity distribution on the CCD for b) δ=0, c) δ=-Γ, d) δ=-2Γ. Solid line gives a fit to diffraction model (see text).